# STRIP-PET: A NOVEL DETECTOR CONCEPT FOR THE TOF-PET SCANNER

P. Moskal, T. Bednarski, P. Białas, M. Ciszewska, E. Czerwiński, A. Heczko, M. Kajetanowicz, Ł. Kapłon, A. Kochanowski, G. Konopka-Cupiał, G. Korcyl, W. Krzemień, K. Łojek, J. Majewski, W. Migdał, M. Molenda, Sz. Niedźwiecki, M. Pałka, Z. Rudy, P. Salabura, M. Silarski, A. Słomski, J. Smyrski, J. Zdebik, M. Zieliński.

Jagiellonian University, Cracow, Poland

**Abstract:**
We briefly present a design of a new PET scanner based on strips of polymer scintillators arranged in a barrel constituting a large acceptance detector. The solution proposed is based on the superior timing properties of the polymer scintillators. The position and time of the reaction of the gamma quanta in the detector material will be determined based on the time of arrival of light signals to the edges of the scintillator strips.

**Introduction:**
At present all commercial PET devices use inorganic scintillator crystals as radiation detectors [1-4]. This is one of the reasons making the PET scanners very expensive. Therefore, we are developing a technology which would allow construction of a PET scanner based on low cost organic scintillators. In the accompanying paper [5] we have argued that disadvantages due to the low detection efficiency and negligible probability for photoelectric effect of organic scintillators can be compensated by large acceptance and improved time resolution achievable with polymer scintillator detectors. In this contribution we present one of the specific solutions referred to as strip-PET which may allow enlargement of diagnostic chambers without an increase of the number of photomultipliers [6].

**Strip-PET:**
The strip-PET test chamber may be formed from strips of organic scintillator. A schematic view of such diagnostic chamber is shown in Fig. 1.

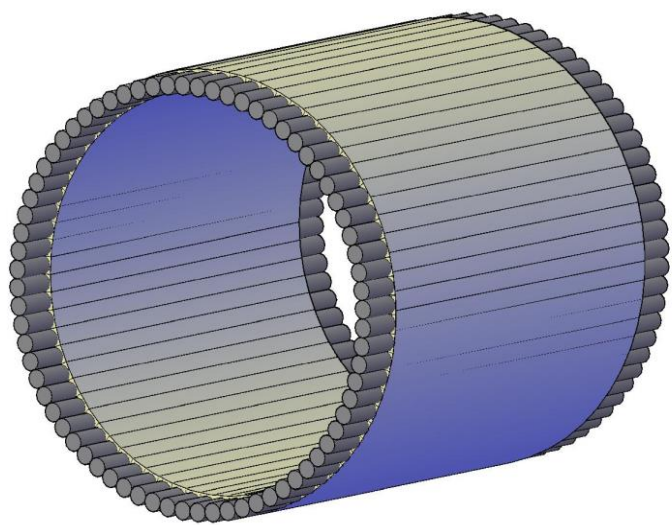

Fig.1. One of the possible arrangements of strips forming a strip-PET diagnostic chamber. Patient would horizontally lie inside the barrel, along scintillator strips.

Light signals from each strip are converted to electrical signals by two photomultipliers placed at opposite edges. The scheme of two detector modules is shown in Fig. 2. The hit position versus the center of the scintillator ($\Delta l$) is determined based on time difference measured on both sides of the scintillation strip. The time at which gamma quantum hits the module can be determined as an arithmetic mean of times measured on both sides of the module. Position ($\Delta x$) along the line of response is determined from time difference between two modules according to the formula shown in Fig. 2. In thin plastic scintillator strips the light signal propagates with velocity of about half of the speed of light in vacuum: c/2. Thus, the resolution of the ($\Delta l$) determination can be expressed as FWHM($\Delta l$) ≈ FWHM($\Delta t$) * c/4, where FWHM($\Delta t$) denotes the resolution of the time difference measurement between photomultipliers on both sides. This translates to FWHM($\Delta t$)/2 of a time resolution of a single module giving FWHM(TOF) = FWHM($\Delta t$)/√2 and as a consequence FWHM($\Delta x$) ≈ FWHM($\Delta t$) * c / 2√2. Hence, this solution requires FWHM($\Delta t$) equal to about 70 ps in order to achieve the resolution of the position determination along the scintillator strip equal to FWHM($\Delta l$) ≈ 0.5 cm. Such resolution of time-of-flight measurement would result in the determination of the annihilation point along the line-of-response with accuracy of FWHM($\Delta x$) ≈ 0.7 cm. The up-to-date detectors do not provide such time precision of the detection of the annihilation gamma quanta. Therefore, the development of the readout electronics and the methods of reconstruction of the time of the signal arrival is at present the main goal of our investigations.

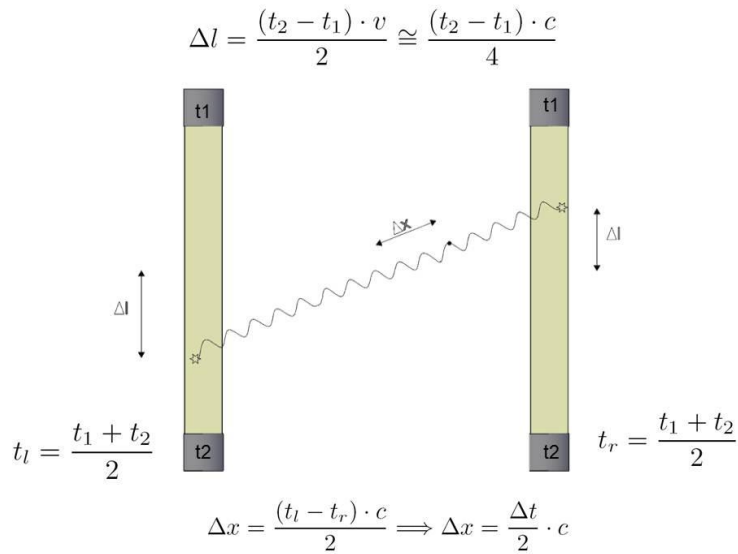

Fig. 2. Schematic view of two modules of the PET detector referred to as strip-PET.

In order to compensate for low density of plastic scintillators several layers of strips could be placed around a patient. It is also worth mentioning that the discussed design of the PET detector allows not only an integration with CT or NMR modality as it is in the case of current scanners, but it also offers new possibilities of combining PET with CT and PET with NMR, so that the same part of the body can be scanned with both methods without moving the patient.

**Acknowledgement:**

Authors acknowledge support of the Foundation for Polish Science and the Polish National Center for Development and Research.